\documentclass[journal]{IEEEtran}
\usepackage{times,amssymb,epsfig}
\usepackage{graphicx}
\usepackage{subfig}
\usepackage{color}
\usepackage{enumerate}
\usepackage{amsmath}
\usepackage{booktabs}
\usepackage{leftidx}
\usepackage{float}
\usepackage{amsmath}
\usepackage{amssymb}
\usepackage[T1]{fontenc}
\usepackage{lineno}
\makeatletter

\newcommand{\Rmnum}[1]{\expandafter\@slowromancap\romannumeral #1@}
\makeatother

\ifCLASSINFOpdf
\else
\fi

\begin{document}
%
\title{Dense People Counting Using IR-UWB Radar \\with a Hybrid Feature Extraction Method}
%
%
%

\author{Xiuzhu Yang, Wenfeng Yin, Lei Li and Lin Zhang \\
Beijing University of Posts and Telecommunications\\
Email: zhanglin@bupt.edu.cn
\thanks{
\emph{Corresponding author: Lin Zhang}
}
}

\maketitle

\begin{abstract}
People counting is one of the hottest issues in sensing applications.
The impulse radio ultra-wideband (IR-UWB) radar has been extensively applied to count people, providing a device-free solution without illumination and privacy concerns. However, performance of current solutions is limited in congested environments due to the superposition and obstruction of signals.
In this letter, a hybrid feature extraction method based on curvelet transform and distance bin is proposed.
2-D radar matrix features are extracted in multiple scales and multiple angles by applying the curvelet transform.
Furthermore, the distance bin is introduced by dividing each row of the matrix into several bins along the propagating distance to select features.
The radar signal dataset in three dense scenarios is constructed, including people randomly walking in the constrained area with densities of 3 and 4 persons per square meter, and queueing with an average distance of 10 centimeters.
The number of people is up to 20 in the dataset.
Four classifiers including decision tree, AdaBoost, random forest and neural network are compared to validate the hybrid features, and random forest performs the highest accuracies of all above 97\% in three dense scenarios. Moreover, to ensure the reliability of the hybrid features, three other features including cluster features, activity features and CNN features are compared. The experimental results reveal that the proposed hybrid feature extraction method exhibits stable performance with significantly superior effectiveness.
\end{abstract}


\section{Introduction}
With the developing requirement for the Internet of Things (IoT) sensing task, estimating the number of people in a monitored area is crucial for sensing applications. Radar systems provide device-free sensing solutions ranging from human detection to activity classification [1], [2].
They leverage radar signals which are reflected and attenuated by human bodies, and infer the valid information by properly analyzing the received signal. 
The impulse radio ultra-wideband (IR-UWB) radar transmits and receives a narrow impulse signal that occupies a wide bandwidth in the frequency domain, with fine delay resolution and excellent penetration. It performs outstanding applications in vital sign monitoring [3], personnel detection [4] and people counting [5]-[7]. Compared with current researches on people counting using vision-based systems [8], the IR-UWB radar doesn't suffer from insufficient illumination and privacy concerns. Moreover, it is a device-free solution without relying on any dedicated or personal device, which is required in other radio-based systems, such as radio frequency identification (RFID), Bluetooth, Zigbee and WiFi [9].

Several studies on people counting using IR-UWB radar are conducted in [5]-[7]. The algorithm in [5] iteratively detects the local maximum of radar signals to count people. In [6], theoretical models of UWB signals are conducted in simulation. [7] proposes an algorithm based on the major clusters, analyzing the distribution of selected amplitudes with the distance and the number of people. These algorithms adequately distinguish multipaths and count people.
However, all of them count each signal separately suffering from ever-changing signals, and the superposition as well as obstruction of signals limit counting performance in congested environments.

In this letter, a hybrid curvelet transform based features-distance bin based features (CTF-DBF) extraction method for dense people counting is proposed.
Firstly, in order to address challenges of rapid variations between signals and superposed multipaths of each signal in congested scenarios, several continuously received signals are regarded as a 2-D radar matrix. Due to the moving continuity and trajectory consistency of people, characteristics of moving people are represented as textures with spatial locality information in the radar matrix. The curvelet transform is applied to extract statistical features in multiple scales with different frequencies as well as multiple angles with diverse moving directions.
Secondly, to extract detailed information and further analyze the superposed and obstructed signal, the distance bin is defined by dividing each signal into several bins along the propagating distance. Characteristics of each distance bin are extracted in an effective way to supplement detailed features for statistical features.
The radar signal dataset comprising three dense scenarios is constructed for 0-20 people randomly walking in the constrained area with densities of 3 and 4 persons per square meter, and at most 15 people queueing with an average distance of 10 centimeters.
\begin{figure*}[!t]
\setlength{\belowcaptionskip}{-0.5cm}
\includegraphics[width=1\linewidth]{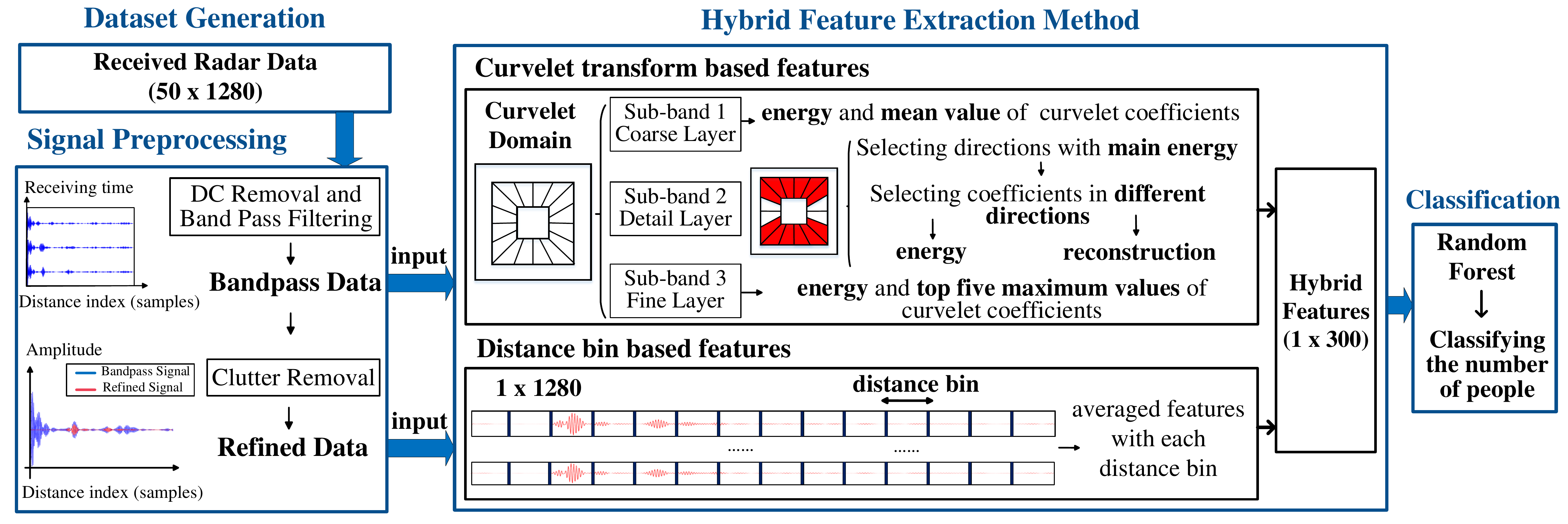}
\captionsetup{font={footnotesize}}
\caption{ Workflow of the people counting system, composed by the dataset generation module, the signal preprocessing module, the proposed hybrid feature extraction method module and the classification module. }
\end{figure*}
With these hybrid features extracted from the dataset, four classifiers including decision tree, AdaBoost, random forest and neural network are compared.
Random forest achieves the highest accuracies in three dense scenarios of all above 97\%.
Furthermore, three other features including cluster features proposed in [7], activity features in [10] and features learnt automatically form LeNet-5 convolutional neural network (CNN) [11] are compared to ensure the reliability of the hybrid features. The experimental results demonstrate the effectiveness and robustness of the hybrid feature extraction method in dense scenarios.

Fig. 1 shows the workflow of the people counting system, composed by the dataset generation module, the signal preprocessing module, the proposed hybrid feature extraction method module and the classification module.
The remainder of this paper is organized as follows. Section \Rmnum{2} describes the dataset generation. The proposed hybrid feature extraction method is discussed in Section \Rmnum{3}. Section \Rmnum{4} presents experimental results and analysis. The conclusions are summarized in Section \Rmnum{5}.
\section{Dataset Generation}
\subsection{Radar System}
In this letter, the IR-UWB radar data from a select number of people in a space is acquired by an NVA-R661 radar module transmitting a narrow pulse with a center frequency of 6.8 GHz, and the bandwidth in -10dB concept of 2.3 GHz. The received radar signals are converted to digital signals, and the sampling frequency is about 39 GHz with the resolution of 0.0039 meter.

The experimental setup is shown in Fig. 2, in which the radar is installed at a height of 1.8 meters, with the detecting range of 5 meters and a central angle of 90 degrees.
To validate the performance of the proposed method, three dense scenarios are considered for radar data collection.
Scenarios 1 and 2 are 0-20 people randomly walking in a constrained area with densities of 3 and 4 persons per square meter respectively.
To maintain the densities unchanged, the activity range of testers is limited in a rectangular region of which the area increases with the increasing number of people, shown in the red area in Fig. 2(a).
Due to the congested environment, the moving speed of people is limited, equal to or less than the normal walking speed.
In scenario 3, at most 15 people stand in a queue with an average distance of 10 centimeters described in Fig. 2(b), and their positions are unchanged. To enhance the reliability of the dataset, 44 testers participate in experiments for acquiring diverse data from different people.

Five seconds of radar data with 200 received signals are recorded for each measurement, and each radar sample is selected independently from the record for 1.25 seconds with 50 received signals.
Radar sample for 1.25 seconds is long enough for curvelet transform based feature extraction, meanwhile counting in every 1.25 seconds is acceptable in a real time system.
Each signal in a radar sample contains 1280 sampling points representing the 5 meters detection range.
3,360 radar samples are generated in scenarios 1 and 2 respectively, with a total of 2,560 samples in scenario 3.
\begin{figure}[!htb]
\setlength{\belowcaptionskip}{-0.6cm}
\centering
\includegraphics[width=2in,height=2.9in]{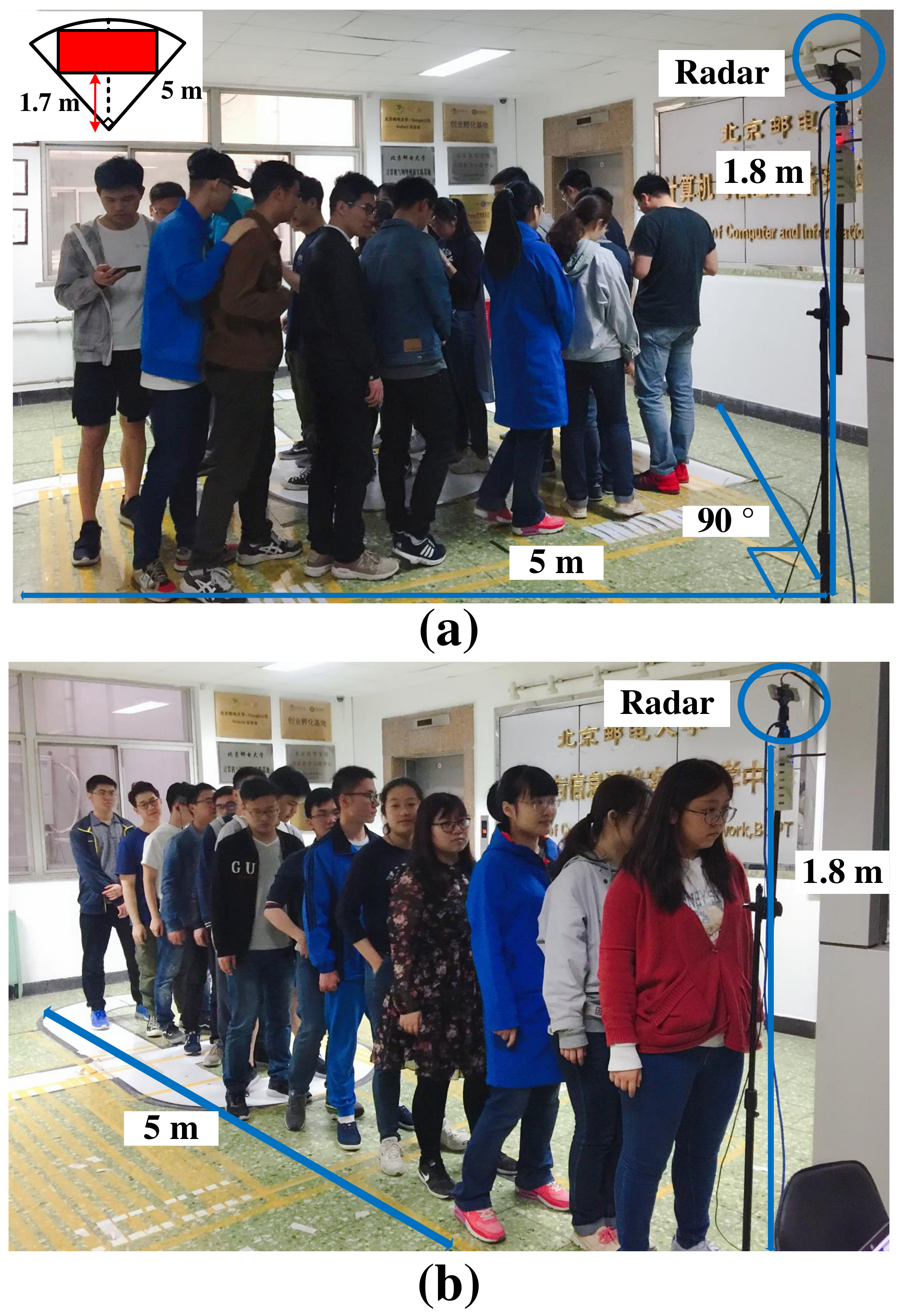}
\captionsetup{font={footnotesize}}
\caption{Experimental setup (a) in the constrained area and (b) in a queue.}
\end{figure}
\subsection{Signal Modeling}
For each received radar sample, the direct current (DC) component is firstly removed, and then a Hamming window is designed as a filter to obtain the bandpass data with frequency from 5.65 GHz to 7.95 GHz, shown in Fig. 1.
The 2-D bandpass data composed by multiple radar signals is described as follows,
\begin{equation}
r(t,x) = p(t,x) + c(t,x) + n(t,x)
\end{equation}
where \textit{t} is the accumulating receiving time representing the time it takes the radar to receive multiple signals, while \textit{x} is the propagating distance of each signal. \textit{p(t,x)} is the target signal reflected from people, while \textit{n(t,x)} is the noise signal. \textit{c(t,x)} represents the clutter, which contains the direct wave from the transmitter to receiver and reflections from the background.
\section{Hybrid Feature Extraction Method}
\subsection{Curvelet Transform based Features}
Fast ever-changing signals and superposed multipaths bring great challenges for dense people counting by amplitudes of each signal. Counting from a single received signal separately is not stable and reliable, therefore the curvelet transform provides statistical features of a radar matrix. Several temporal continuously received signals are considered as a 2-D radar matrix to avoid contingency caused by the single signal. Furthurmore, considering the moving continuity and trajectory consistency, trajectories of people are presented as textures with spatial locality information in the matrix. Superposed multipaths show stronger textures and can be obviously observed with the curvelet transform. In addition, the curvelet transform provides a multi-scale and multi-orientation decomposition for the 2-D radar matrix to adequately represent texture and edge information with curve-like features [12], providing information on signal strength and moving direction of people. The definition of discrete Curvelet transform is given as follows,
\begin{equation}
C(j,l,k) =  \int \hat{f}(\omega)\hat{U_j}(S^{-1}_{\theta_l}\omega)e^{i<S^{-T}_{\theta_l}b,\omega>}d\omega\
\end{equation}
where $j$, $l$ and $k$ are the parameters of the scale, the direction and the position.
$f$ represents the input radar data in the Cartesian coordinate system. $U_j$ is the frequency window for each scale $j$, and $S_{\theta_l}$ is the shear matrix with orientation $\theta_l$ defined as
\begin{equation}
S_{\theta_l} :=
\left(
  \begin{array}{ccc}
    1 & 0 \\
    -tan\theta_l & 1\\
  \end{array}
\right)
\end{equation}
where superscript $T$ represents the transpose of the matrix. $b$ is defined as $b :=  (k_1\cdot2^{-j}, k_2\cdot2^{-j/2})$, where the sequence of translation parameters $k=(k_1,k_2)\in Z^2$.

When people are queueing in a line or remaining still, their positions are unchanged in a period of time, forming straight-like lines in the 2-D image. In this case, signals from people with smaller variances are easily to be mistaken as clutters reflected from background, thus removing clutters will eliminate significant information as well. To fully extract statistical features from radar matrix without losing any useful information, the bandpass data matrix without clutter removal is decomposed by the curvelet transform, shown in Fig. 1. Each bandpass data with 50 continuously received signals and 1280 sampling points in each signal is decomposed into a coarse layer, a detail layer and a fine layer representing different scales.
To characterize the signal matrix in all scales, features with three layers are extracted in the curvelet domain.
The coarse layer formed by low frequency coefficients shows the general characteristic and the tendency information of the signal matrix, thus the mean value and energy of the coarse layer are extracted to generally describe the radar data. The fine layer contains high frequency coefficients, representing the finer edge information, which is usually represented by the maximum value. Therefore the top five maximum values as well as the energy of the fine layer are extracted.
\begin{figure}[!t]
\setlength{\belowcaptionskip}{-0.2cm}
\centering
\includegraphics[width=3.2in,height=1.6in]{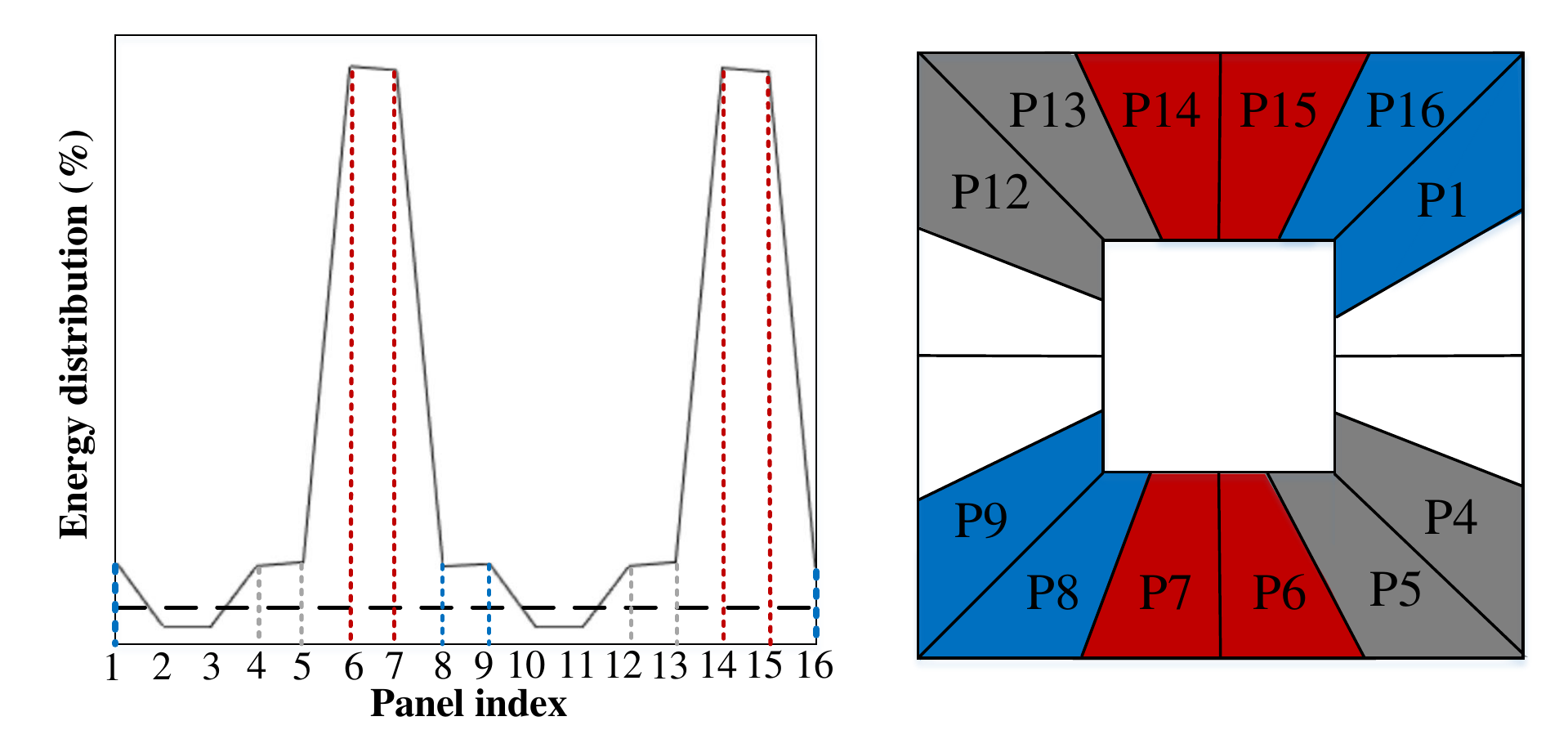}
\captionsetup{font={footnotesize}}
\caption{Energy distribution of 16 coefficient matrices in the detail layer with corresponding panels in the radar matrix. The red, blue and gray dashed lines parallel to the vertical axis are the energy of corresponding panels marked in the same colour.}
\end{figure}
\begin{figure}[!t]
\setlength{\belowcaptionskip}{-0.7cm}
\centering
\includegraphics[width=2.3in,height=1.5in]{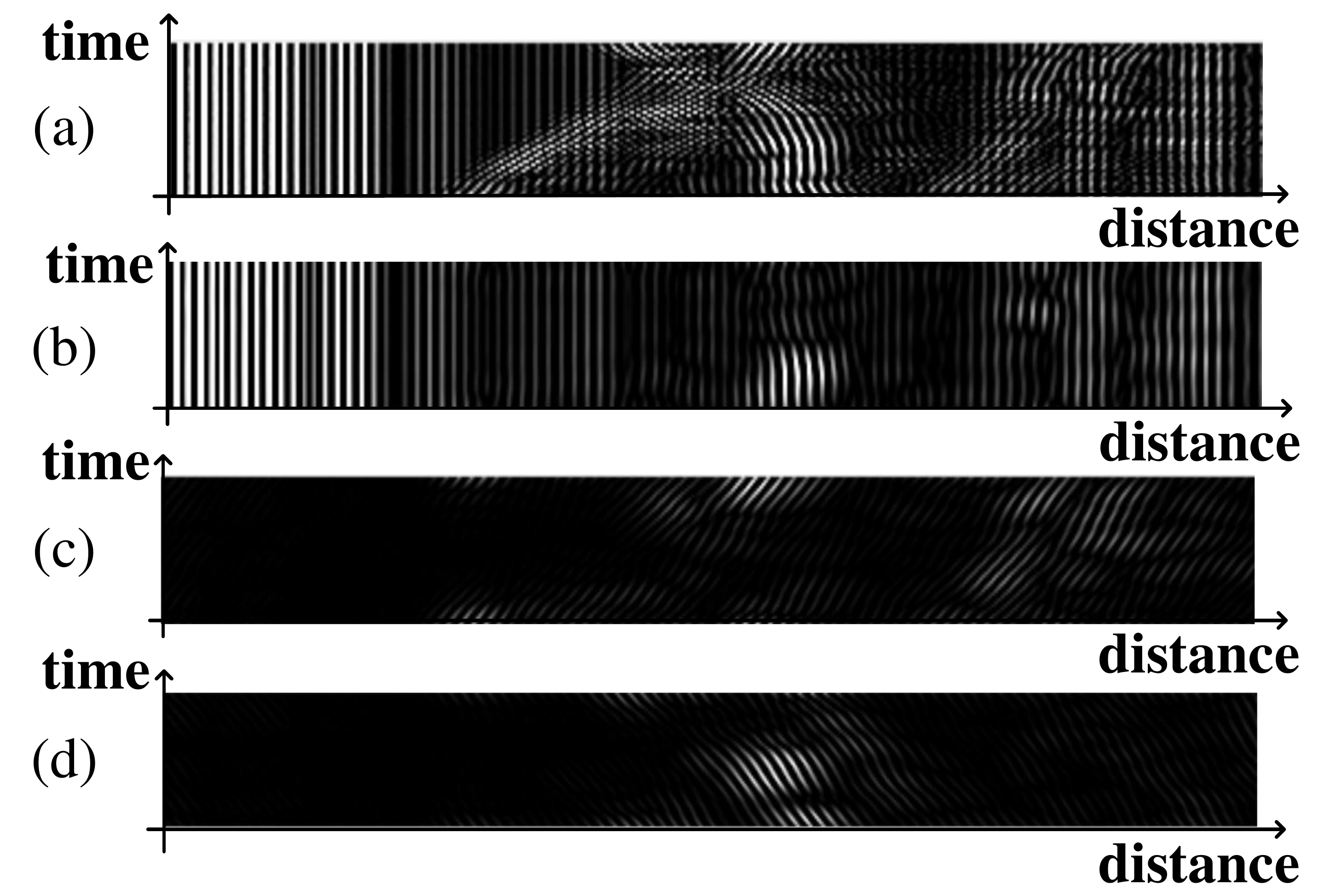}
\captionsetup{font={footnotesize}}
\caption{The radar matrix for (a) the bandpass data, and the reconstructed data by (b) the $90^{\circ}$ vertical coefficients (c) the $45^{\circ}$ diagonal coefficients (d) the $135^{\circ}$ diagonal coefficients of the detail layer.}
\end{figure}

The detail layer with high frequency coefficients is divided into 16 directions.
Panels of the radar matrix in Fig. 3 are arranged in the clockwise direction, and each angular panel occupies $22.5^{\circ}$.
Coefficients in each panel represent signals on corresponding moving directions in trajectories of people. In order to increase the stability and reliability of extracted features, the blue dashed line parallel to the horizontal axis in Fig. 3 cuts the energy, and coefficients with too low energy are removed.
Fig. 4 shows the reconstructed signals by picking up coefficients in corresponding directions, which are presented as textures of moving trajectories in the grey-scale maps.
Panel 1, 16, 8 and 9 in blue are selected as the $45^{\circ}$ direction shown in Fig. 4(c), representing people moving further away from the radar in 1.25 seconds, while panel 4, 5, 12 and 13 in gray are extracted for the $135^{\circ}$ direction in Fig. 4(d) representing people moving closer to the radar during this time. Panel 6, 7, 14 and 15 in red representing the $90^{\circ}$ direction in Fig. 4(b) occupy most of the energy due to the static clutter, as well as reflections from people on the spot.
Considering superposed multipaths for stronger textures, the energy for each direction are calculated in the curvelet domain and in reconstructed signals respectively to represent the comprehensive information of people in the corresponding direction.
\subsection{Distance Bin based Features}
To get detailed features for complementing curvelet transform based statistical features and further analyze the superposed and obstructed signal in dense people counting, several features are extracted from each signal.
Clutter signals are removed firstly using a running average based method [5] to analyze the valid signals reflected from people, and feature extraction is operated on the refined data in red, shown in Fig. 1.

Due to the high sampling rate in received radar signals, the redundant information contained in these samples will cause over-fitting. To select the representative information for the number of people and reduce over-fitting, each signal with 1280 sampling points is divided into several bins along the propagating distance, with each length of \textit{S$_{d}$}.

\begin{figure}[!htb]
\setlength{\belowcaptionskip}{-0.2cm}
\centering
\includegraphics[width=3in,height=1.2in]{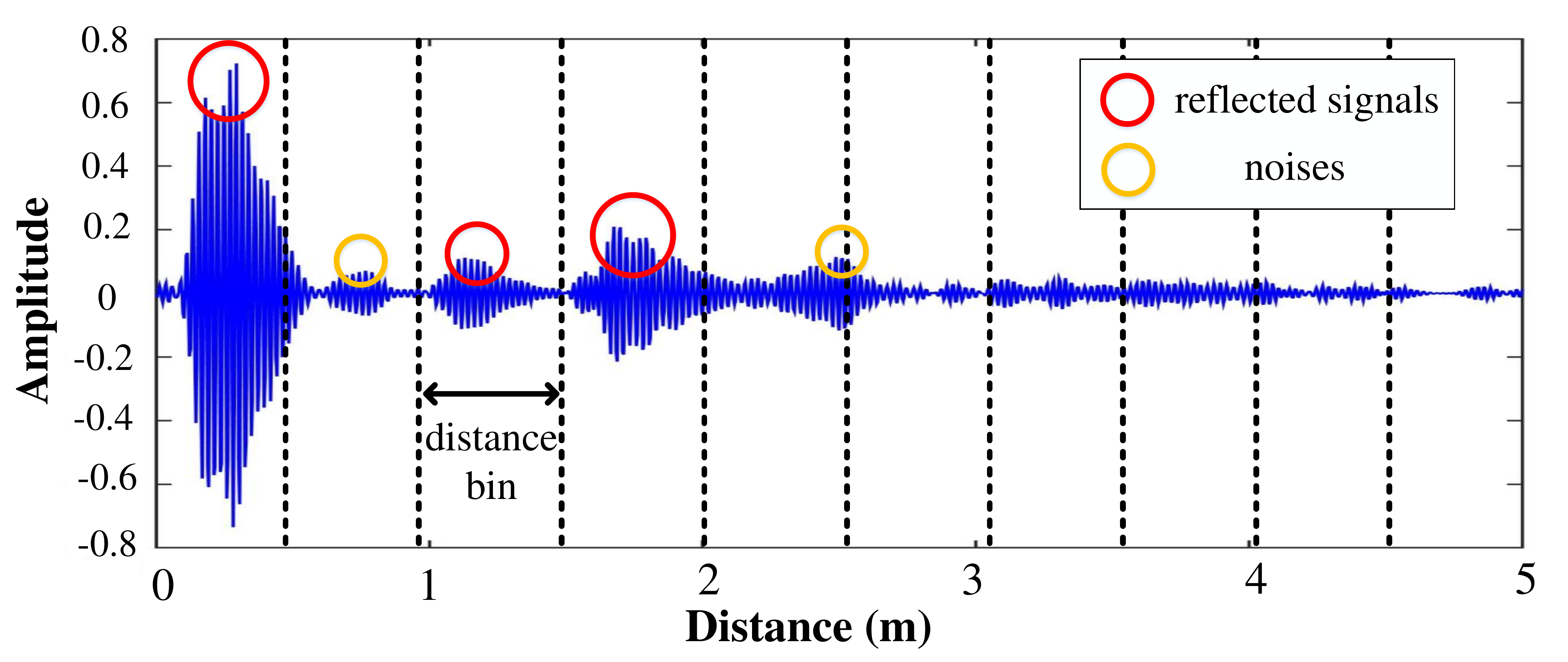}
\captionsetup{font={footnotesize}}
\caption{ Refined radar signal for 4 people in a queue.}
\end{figure}

The maximum amplitude in each distance bin is selected as a feature to represent a candidate point for the presence of people.
However, the number of local maximum amplitudes can not represent the number of people when they stand closely. As shown in Fig. 5, the first red circle is obviously detected for the presence of 1 person, but there are 2 persons standing closely. In this case, it is impossible to distinguish the number of people from the amplitude due to the multipaths from different people. But the energy of different people are superposed and distinct significantly in a distance bin, therefore the energy is calculated as the complementary feature by squaring sampled signals and integrating them over each bin.

Since that the transmitted power of UWB radar is limited and the relatively high noise is accumulated over its wideband. The signal-to-noise ratio (SNR) is low, thus considering the energy of each bin also takes noises into calculation and meanwhile amplifies them. In dense queueing counting, the reflected signal from obstructed people is severely attenuated and comparable with the noise in observation, as Fig. 5 describes. Amplitudes marked by two orange circles are noises in the environment, but their amplitudes are comparable with reflected signals from people, marked in the second and third red circles.
Therefore, the noise is removed using the hard threshold analysis of the curvelet transform [13]. Then the energy and maximum amplitude of denoised signals are also extracted as features.

The length of each distance bin $S_d$ is of great importance in identifying the detailed features. To discriminate each person from the superposed multipath signals in congested environments, $S_d$ should be smaller than a certain physical parameter, for example, the height or the shoulder width of a person [7].
In order to obtain sufficient features in different scales to better describe the detailed information, the distance bin is chosen for 125, 250 and 500 millimeters respectively.
In this letter, the spatial resolution of the radar system for the minimum distinguishable distance between two adjacent sampling points is 3.9 millimeters. Therefore, the number of sampling points in each distance bin is set as 32, 64 and 128 respectively. To maintain the length of the distance bin, radar system with higher spatial resolution needs more sampling points in each distance bin, while radar system with lower resolution needs less sampling points.
For the radar matrix with multiple signals, these features are averaged respectively.

The distance bin based features are then directly combined with the curvelet transform based features as the hybrid features, which are defined more clearly in Table \Rmnum{1}.
\begin{table}[!htb]
\newcommand{\tabincell}[2]{\begin{tabular}{@{}#1@{}}#2\end{tabular}}
\centering
\caption{Hybrid CTF-DBF features}
\begin{tabular}{ll}
Terms & Definition\\
\hline
\tabincell{l}{Features of coarse layer \\in the curvelet domain}& \tabincell{l}{Mean and energy of curvelet \\coefficients in the coarse layer}  \\
\hline
\tabincell{l}{Features of fine layer \\in the curvelet domain}& \tabincell{l}{Top five maximum values and energy \\ of curvelet coefficients in the fine layer}\\  
\hline
\tabincell{l}{Features of detail layer \\in the curvelet domain}& \tabincell{l}{Energy of the $90^{\circ}$ vertical coefficients, \\$45^{\circ}$ diagonal coefficients and $135^{\circ}$  \\ diagonal coefficients}\\
\hline
\tabincell{l}{Features of detail layer \\in the reconstructed signal}& \tabincell{l}{Energy of the reconstructed signal with \\the $90^{\circ}$ vertical coefficients, $45^{\circ}$   \\diagonal coefficients and $135^{\circ}$ \\diagonal coefficients}\\
\hline
\hline
\tabincell{l}{Number of sampling points \\in a bin \emph{$S_d$}} & \tabincell{l}{The number of sampling points in \\a distance bin, with domain\{32, 64, 128\}}\\
\hline
Maximum Amplitude \emph{$A_k$ / $A_d$} & \tabincell{l}{The maximum amplitude of each \\distance bin for signals with and \\without noises with corresponding \emph{$S_d$}}\\
\hline
Energy \emph{$E_{k}$ /  $E_{d}$} & \tabincell{l}{The energy of each distance bin \\for signals with and without noises, \\with corresponding \emph{$S_d$}}\\
\hline
\end{tabular}
\end{table}
\vspace{-0.3cm}

\section{Experimental Results}
\subsection{Performance on Different Classifiers}
The hybrid CTF-DBF feature samples extracted from the dataset constructed in this letter with size 1 $\times$ 300 are used as input for classification.
To verify the effectiveness of the hybrid feature extraction method, four selected classifiers are compared, including decision tree, AdaBoost, random forest and neural network.
The decision tree is a rooted tree structure to divide the cases into two subtrees in each node.
In this letter, the random forest classifier constructs 200 decision trees to increase the classification ability. The AdaBoost consists 50 base estimators, with the SAMME.R classification algorithm and a linear loss function. The neural network has three hidden layer with 100, 200 and 100 neurons respectively as well as a ReLU activation function, and is optimized by Adam algorithm.

The feature samples are divided into the training set and the testing set.
80\% randomly chosen samples (2688 samples in scenarios 1 and 2 respectively, and 2048 samples in scenario 3) are used as the training set for the supervised training on each classifier, and the remaining 20\% feature samples are as the testing set to test the classifier and calculate the classification error. The calculation for each classifier is repeated for 20 times with randomly chosen training data, and the average accuracy, precision, recall and F1 score [14] are computed, shown in Table \Rmnum{2}, \Rmnum{3} and \Rmnum{4}.

\begin{table}[!htb]
\centering
\caption{Classification performance comparison of different classifiers for 0-20 people randomly walking in the constrained area with 3 persons per square meter.}
\begin{tabular}{l c c c c}
\hline
 & \multicolumn{1}{c}{Accuracy} & \multicolumn{1}{c}{Precision} & \multicolumn{1}{c}{Recall}& \multicolumn{1}{c}{F1}\\

\hline
Decision Tree &76.7\%&87.6\%&87.0\% &87.3\%\\
AdaBoost &80.5\%&90.9\%&91.9\% &91.4\%\\
\textbf{Random Forest} &\textbf{97.6\%}&\textbf{97.5\%}&\textbf{99.5\%}&\textbf{98.5\%}\\
Neural Network &95.1\%&97.8\%&95.3\%&96.5\% \\
\hline
\end{tabular}
\end{table}

\begin{table}[!htb]
\centering
\caption{Classification performance comparison of different classifiers for 0-20 people randomly walking in the constrained area with 4 persons per square meter.}
\begin{tabular}{l c c c c}
\hline
 & \multicolumn{1}{c}{Accuracy} & \multicolumn{1}{c}{Precision} & \multicolumn{1}{c}{Recall}& \multicolumn{1}{c}{F1}\\

\hline
Decision Tree &76.6\%&87.6\%&87.0\% &87.3\%\\
AdaBoost &81.2\%&90.9\%&91.9\% &91.4\%\\
\textbf{Random Forest} &\textbf{97.5\%}&\textbf{97.4\%}&\textbf{99.4\%}&\textbf{98.4\%}\\
Neural Network &94.9\%&98.4\%&95.0\%&96.7\% \\
\hline
\end{tabular}
\end{table}

\begin{table}[!htb]
\centering
\caption{Classification performance comparison of different classifiers for 0-15 people in the queue.}
\begin{tabular}{l c c c c}
\hline
 & \multicolumn{1}{c}{Accuracy} & \multicolumn{1}{c}{Precision} & \multicolumn{1}{c}{Recall}& \multicolumn{1}{c}{F1}\\

\hline
Decision Tree &83.8\%&88.9\%&90.0\% &89.4\%\\
AdaBoost &87.3\%&93.1\%&92.5\% &92.8\%\\
\textbf{Random Forest} &\textbf{98.7\%}&\textbf{99.4\%}&\textbf{100.0\%}&\textbf{99.7\%}\\
Neural Network &97.8\%&99.5\%&99.2\%&99.4\% \\
\hline
\end{tabular}
\end{table}

As Table \Rmnum{2}, \Rmnum{3} and \Rmnum{4} describe, the accuracies of random forest and neural network in three dense scenarios are all above 94\%, proving the effectiveness and robustness of the hybrid features in dense people counting.
Random forest performs the highest accuracies of 97.6\% and 97.5\% in the constrained area with 3 and 4 persons per square meter respectively, and has a best mean accuracy (98.7\%) in the queueing counting. The accuracy, precision, recall and F1 are all above 97\% on random forest, demonstrating the extremely satisfactory classification performance.
The classification accuracies on 3 persons per square meter are similar to that of 4 persons per square meter on four classifiers, indicating that the hybrid features extraction method is robust and insensitive to the dense environments.

\subsection{Performance Comparison with Other Features}
In order to verify the superiority of the proposed hybrid features, three other features are used for comparison, including the cluster features proposed in [7], the activity features in [10] and the features learnt automatically from LeNet-5 convolutional neural network (CNN) [11].
The cluster features are composed by the detected amplitudes and distances of the corresponding cluster, with the size of 1$\times$1280.
The activity features consist "activity event" and "activity duration", extracted with the size of 1$\times$2957.
The CNN features are extracted by using the LeNet-5 neural network, which is trained on radar samples for 20 epoches. Features are extracted from the fully connected layer with the size of 1$\times$500.
The comparison results are conducted with random forest in three dense scenarios, described in Fig. 6.

\begin{figure}[!htb]
\setlength{\belowcaptionskip}{-0.4cm}
\centering
\includegraphics[width=3.3in,height=2.2in]{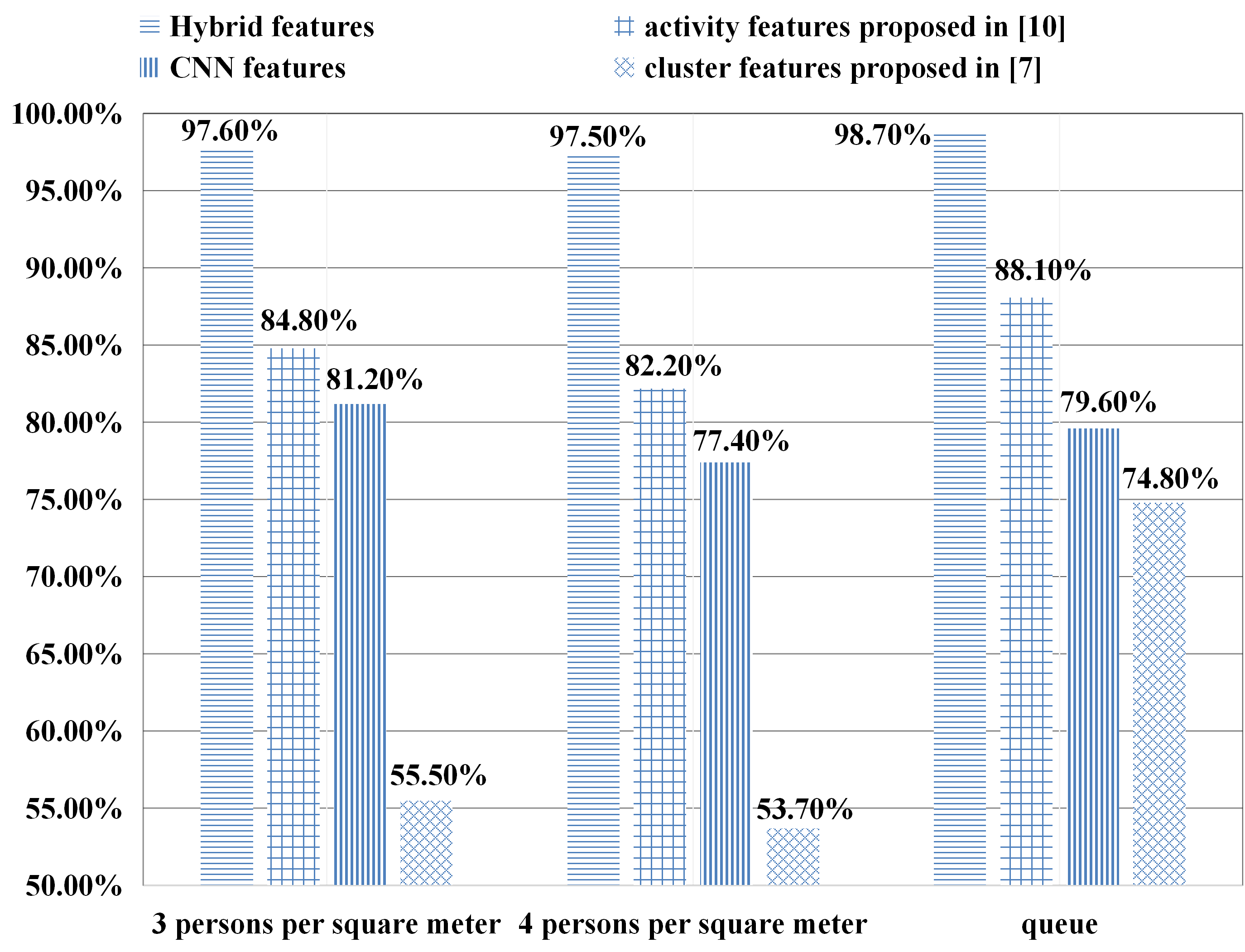}
\captionsetup{font={footnotesize}}
\caption{ Classification accuracies comparison on different features in three dense scenarios.}
\end{figure}

As illustrated, the classification accuracies of proposed hybrid features are obviously better than those of three other features, and have a distinct advantage of stable performance in three dense scenarios, especially in more complex scene of 4 persons per square meter.
Results reveal that the proposed hybrid features are significantly superior to other features in dense people counting.
\section{Conclusion}
In this letter, a hybrid CTF-DBF feature extraction method for dense people counting with IR-UWB radar is proposed. Features with multiple scales and multiple orientations are extracted from the radar matrix by applying the curvelet transform. Moreover, the distance bin is introduced to divide each row of the matrix into several bins along the propagating distance to select features as supplementary information. The dataset in three dense scenarios is constructed, and four classifiers are compared. Counting accuracies are all above 97\% with random forest. Moreover, three other features are compared to verify the superiority of the hybrid features. Comparison results prove the effectiveness and robustness of the proposed method in dense scenarios. In the future work, more radar samples will be collected in more complex scenarios to validate the robustness of the proposed method.

\ifCLASSOPTIONcaptionsoff
  \newpage
\fi

%








\end{document}